\documentclass[a4paper,fleqn,usenatbib,useAMS]{mnras}

\usepackage{graphicx}	
\usepackage{amsmath}	
\usepackage{amssymb}	
\usepackage{multicol}        
\usepackage{bm}		
\usepackage{pdflscape}	

\usepackage[T1]{fontenc}
\usepackage{ae,aecompl}

\title[{\it XMM-Newton} Observations of NGC\,6888]{X-ray emission from the Wolf-Rayet bubble NGC\,6888. \\ II. {\it XMM-Newton} EPIC observations}
\author[J.A.\,Toal\'{a} et al.]{J.A.\,Toal\'{a}\thanks{E-mail: toala@iaa.es}$^{1}$, M.A.\,Guerrero$^{1}$, Y.-H.\,Chu$^{2}$, S.J.\,Arthur$^{3}$, D. Tafoya$^{3}$ and R.A.\,Gruendl$^{4}$\\
$^{1}$Instituto de Astrof\'{i}sica de Andaluc\'{i}a, IAA-CSIC, Glorieta de la Astronom\'{i}a s/n, 18008 Granada, Spain\\
$^{2}$Institute of Astronomy and Astrophysics, Academia Sinica (ASIAA), Taipei 10617, Taiwan\\ 
$^{3}$Instituto de Radioastronom\'{i}a y Astrof\'{i}sica, UNAM, Campus Morelia, Apartado Postal 3-72,58090, Morelia, Michoac\'{a}n, Mexico\\
$^{4}$Department of Astronomy, University of Illinois, 1002 West Green Street, Urbana, IL 61801, USA
}

\pubyear{2015}

\begin{document}
\label{firstpage}
\pagerange{\pageref{firstpage}--\pageref{lastpage}}
\maketitle

\begin{abstract}
  We present deep \emph{XMM-Newton} EPIC observations of the
  Wolf-Rayet (WR) bubble NGC\,6888 around the star WR\,136. The
  complete X-ray mapping of the nebula confirms the distribution of
  the hot gas in three maxima spatially associated with the caps and
  northwest blowout hinted at by previous \emph{Chandra}
  observations. The global X-ray emission is well described by a
  two-temperature optically thin plasma model
  ($T_{1}$=1.4$\times$10$^{6}$~K, $T_{2}$=8.2$\times$10$^{6}$~K) with
  a luminosity of $L_{\mathrm{X}}$=7.8$\times$10$^{33}$~erg~s$^{-1}$
  in the 0.3--1.5~keV energy range. The rms electron density of the
  X-ray-emitting gas is estimated to be
  $n_\mathrm{e}$=0.4~cm$^{-3}$. The high-quality observations
  presented here reveal spectral variations within different regions
  in NGC\,6888, which allowed us for the first time to detect
  temperature and/or nitrogen abundance inhomogeneities in the hot gas
  inside a WR nebula. One possible explanation for such spectral
  variations is that the mixing of material from the outer nebula into
  the hot bubble is less efficient around the caps than in other
  nebular regions.

\end{abstract}





\section{INTRODUCTION}
\label{sec:intro}

The process of formation and evolution of Wolf-Rayet (WR) bubbles is
directly related to the evolution of massive stars that started their
lives with initial masses $\gtrsim25$~M$_{\odot}$ \citep[e.g.,][and
references therein]{Ekstrom2012,Georgy2012}. 
The formation scenario of WR bubbles was refined in the 90s by the works 
presented by \citet{GS1995} and \citet{GS1996a,GS1996b}. 
In this scenario the star evolves from the
main-sequence (MS) phase to a red supergiant (RSG) or luminous blue
variable (LBV). In the RSG/LBV phase the star loses its envelope into
the interstellar medium (ISM) via a copious, slow and dense wind
($v_\infty$=10--100~km~s$^{-1}$ and
$\dot{M}=$10$^{-4}$--10$^{-3}$~M$_{\odot}$~yr$^{-1}$). When the star
evolves to the WR phase, it develops a powerful wind
($v_\infty$=1000--2500~km~s$^{-1}$ and $\dot{M} \approx
$10$^{-5}$~M$_{\odot}$~yr$^{-1}$) and a strong ionizing photon flux
that interact with the circumstellar material (CSM), creating the WR
nebula. 

The wind-wind interaction described above is thought to produce the
so-called adiabatically-shocked hot bubble. The expected plasma
temperature of this hot bubble is directly related to the terminal
wind velocity as $kT=3 \mu m_\mathrm{H} v_{\infty}^{2}/16$, where $k$
is the Bolztmann constant and $\mu$ is the mean particle mass for
fully ionized gas \citep[see][]{Dyson1997}. That is, for a typical WR
wind velocity, the expected temperature of the hot bubble would be
$T$=10$^{7}$--10$^{8}$~K.

The nebula NGC\,6888 around WR\,136 is the most studied WR bubble in
X-rays.  The distribution of the X-ray emission derived from early
observations by \textit{Einstein}, \textit{ROSAT}, and \textit{ASCA}
is interpreted to be associated with the clumpy H$\alpha$ distribution
\citep{B1988,Wrigge1994,Wrigge2002,Wrigge2005}. Recently,
\textit{Suzaku} and \textit{Chandra} observations have been used to
obtain partial maps of the X-ray emission within this nebula
\citep{Zhekov2011,Toala2014}. The \textit{Suzaku} observations of the
northern and southern caps seen in optical H$\alpha$ images
\citep[see][]{Gruendl2000,Stock2010} imply that the X-ray emission in
NGC\,6888 could be described by a two-temperature plasma model with
temperatures $T_{1}<$5$\times$10$^{6}$~K and a small contribution from
a much hotter plasma component $T_{2}>$2$\times$10$^{7}$~K
\citep{Zhekov2011}.  The high-spatial resolution of the
\textit{Chandra} ACIS observations presented by \citet{Toala2014}
showed that the higher temperature component in the \textit{Suzaku}
spectrum was due to contamination by point-like sources projected
along the line of sight of the nebula not resolved by those
observations.  They concluded that a two-temperature plasma model of
$T_{1}\sim$1.4$\times$10$^{6}$~K and $T_{2}\sim$7.4$\times$10$^{6}$~K
described more accurately the spectral properties of the hot gas
inside NGC\,6888.  Furthermore, \citet{Toala2014} argued that there
was an additional component in the spatial distribution of the
X-ray-emitting gas in NGC\,6888 toward the northwest blowout observed
in [O\,{\sc iii}] optical narrowband and mid-infrared images.  This
feature is reminicent of the blowout seen in the WR bubble S\,308
\citep{Chu2003,Toala2012} and that reported recently by
\citet{Toala2015} for NGC\,2359.

In this paper we present new \textit{XMM-Newton} EPIC observations of
NGC\,6888, which confirm the presence of the component in the diffuse
X-ray emission toward the northwest blowout.  We also report for the
first time significant spectral variations among different regions of
NGC\,6888 (the caps, central regions, and blowout).  The paper is
organized as follows: in Section~2 we present the \textit{XMM-Newton}
observations, Sec.~3 and Sec.~4 describe the distribution of the
diffuse X-ray emission and the spectral properties, respectively.  We
discuss our findings in Sec.~5 and summarize the main results in
Sec.~6.

\section{OBSERVATIONS}

The WR bubble NGC\,6888 was observed by \textit{XMM-Newton} on 2014
April 5 (Observation ID 0721570101; PI: J.A.\ Toal\'{a}) using the
European Photon Imaging Camera (EPIC) in the extended full-frame mode
with the medium optical filter for a total exposure time of 75.9~ks. 
The corresponding net exposure times for the EPIC-pn, EPIC-MOS1, and 
EPIC-MOS2 cameras were 61.1 ks, 73.8 ks, and 73.8 ks, respectively. 
The observations were processed using the \textit{XMM-Newton} Science
Analysis Software (SAS) version 13.5 and the Calibration Access Layer
available on 2014 April 14.  
The Observation Data Files (ODF) were reprocessed using the SAS tasks 
\textit{epproc} and \textit{emproc} to produce the corresponding event 
files.

The {\it XMM-Newton} EPIC observations of NGC\,6888 were analysed
following a procedure similar to that described in \citet{Toala2015}
for the case of NGC\,2359. We first present the data analysis, making
use of the \textit{XMM-Newton} Extended Source Analysis Software
package \citep[{\sc xmm-esas};
see][]{Kuntz2008,Snowden2004,Snowden2008} in order to achieve a
detailed description of the distribution of the X-ray-emitting gas
(see Section~3). Secondly, we present the X-ray spectra of the diffuse
emission making use of the SAS tasks {\it evselect}, {\it arfgen}, and
{\it rmfgen} and produce the associated calibration matrices as
described in the SAS threads (see Section~4). The resulting
background-subtracted spectra were used to study the physical
parameters of the X-ray-emitting gas in NGC\,6888.

\begin{figure*}
\begin{center}
\includegraphics[angle=0,width=1.0\linewidth]{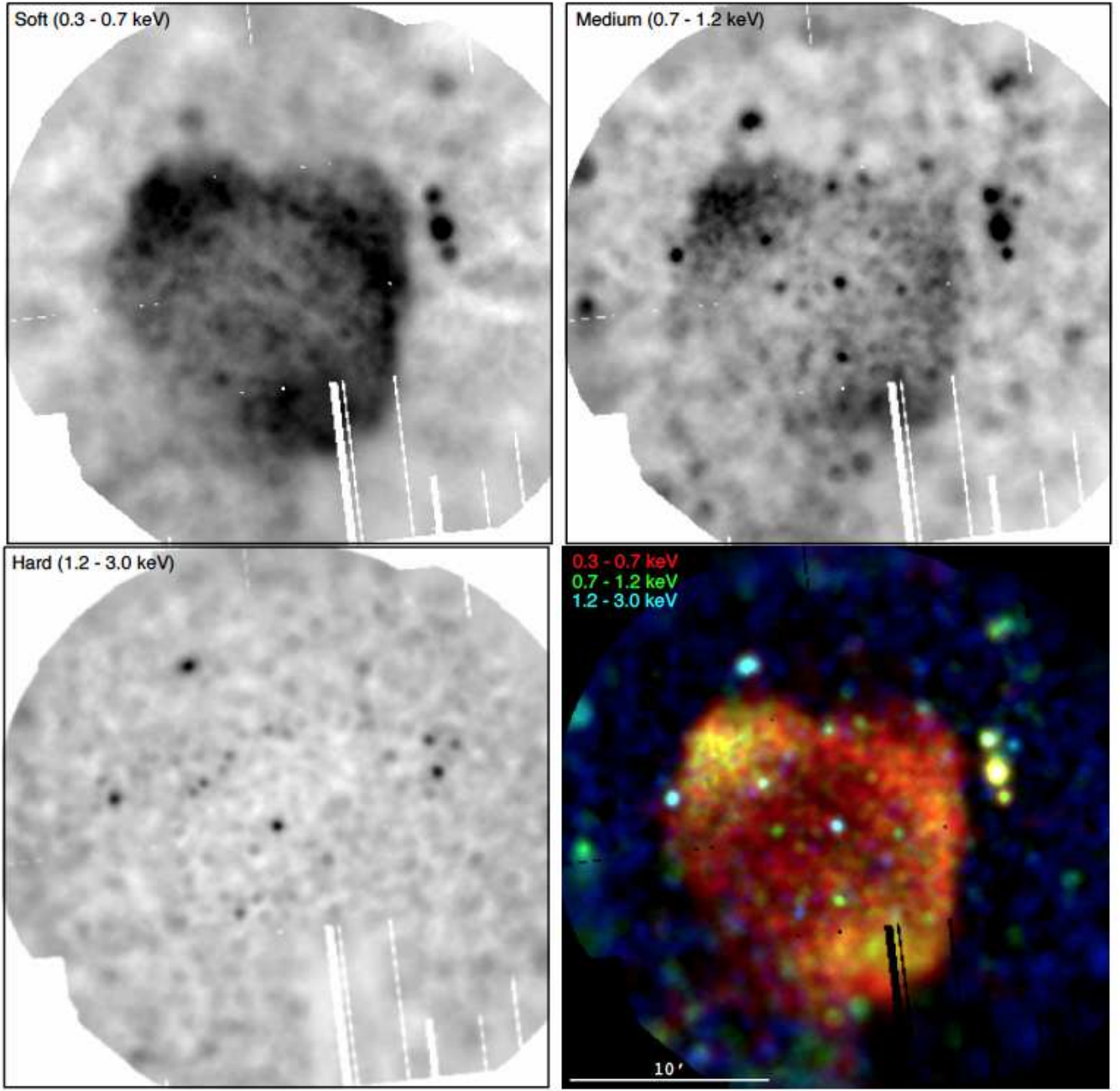}
\label{fig:ESAS}
\caption{
\textit{XMM-Newton} EPIC (MOS1+MOS1+pn) exposure-corrected X-ray images 
in three different energy bands of the field of view of NGC\,6888. 
The images in these three energy bands have been combined to produce the 
colour-composite picture shown in the bottom-right panel.  
The central star in NGC\,6888 (WR\,136) is located at the center of 
each image. 
Note that the point sources have not been excised from the observations. 
North is up, east to the left.  
}
\end{center}
\end{figure*}

\section{SPATIAL DISTRIBUTION OF THE DIFFUSE X-RAY EMISSION}

We followed the Snowden \& Kuntz cookbook for the analysis of the
\textit{XMM-Newton} EPIC observations of extended objects and diffuse
background Version 5.9\footnote{ The online version of the {\sc
    XMM-ESAS} cookbook can be found at
  ftp://xmm.esac.esa.int/pub/xmm-esas/xmm-esas.pdf}.  The associated
Current Calibration Files for the {\sc esas} task can be obtained from
{\small
  ftp://xmm.esac.esa.int/pub/ccf/constituents/extras/esas\_caldb/}.
These calibration files were used to successfully remove the
contribution from the astrophysical background, the soft proton
background, and solar wind charge-exchange reactions, which have
important contributions at energies $<$1.5~keV.  After the restrictive
selection criteria of the {\sc esas} tasks, the final net exposure
times of the pn, MOS1, and MOS2 cameras are 13.2 ks, 39.4 ks, and
43.3~ks, respectively.

Taking into account the spectral shape of the diffuse X-ray emission
reported from {\it Chandra} and {\it Suzaku} observations, we have
created EPIC images in the energy bands 0.3--0.7 keV, 0.7--1.2 keV,
and 1.2--3.0 keV which we will label as soft, medium, and hard bands.
Individual EPIC-pn, EPIC-MOS1, and EPIC-MOS2 images were extracted,
merged together, and corrected for exposure maps.  The final
exposure-map-corrected, background-subtracted EPIC images are
presented in Figure~1 as well as a color-composite X-ray picture of
the three bands.  Each figure has been adaptively smoothed using the
{\sc esas} task \textit{adapt} requesting 100~counts under the
smoothing kernel for the soft and medium bands and 50~counts for the
hard band of the original images.  Sadly, on top of the MOS1 CCD\,3
and CCD\,6, which are no longer functional, CCD\,4 had to be excised
also during the data analysis. As stated in the {\sc xmm-esas}
cookbook, this CCD has a tendency to increased levels of emission at
low energies toward its right side.  As a result, some regions in the
final images presented in Figure~1 show a number of gaps.

Figure~1 shows the presence of diffuse X-ray emission toward NGC\,6888
as reported by previous works \citep[e.g.,][and references
therein]{Zhekov2011,Toala2014} as well as a number of point-like
sources in the field of view of the observations including the
progenitor star, WR\,136.  Furthermore, it is clear that the X-ray
emission has three maxima associated with the northern and southern
caps, and with the northwest blowout detected in [O\,{\sc iii}]
optical narrowband and {\it Spitzer} MIPS 24~$\mu$m images
\citep[see][]{Gv2010,MesaDelgado2014}.

The diffuse X-ray emission of NGC\,6888 shows remarkable spectral
variations across the nebula.  The bulk emission from the caps and
blowout is found in the soft band, with a significant contribution in
the medium band.  On the other hand, while the central and southeast
regions of the nebula are also strong in the soft band, their emission
in the medium band is neglible. Furthermore, none of these regions has
appreciable emission in the hard band.
The differing spatial distribution of the X-ray emission in the soft
and medium energy bands hints at spectral differences between regions
within the nebula never reported by previous X-ray studies of
NGC\,6888. Note that no appreciable diffuse X-ray emission is detected
in the hard band.

\begin{figure*}
\begin{center}
\includegraphics[angle=0,width=1\linewidth]{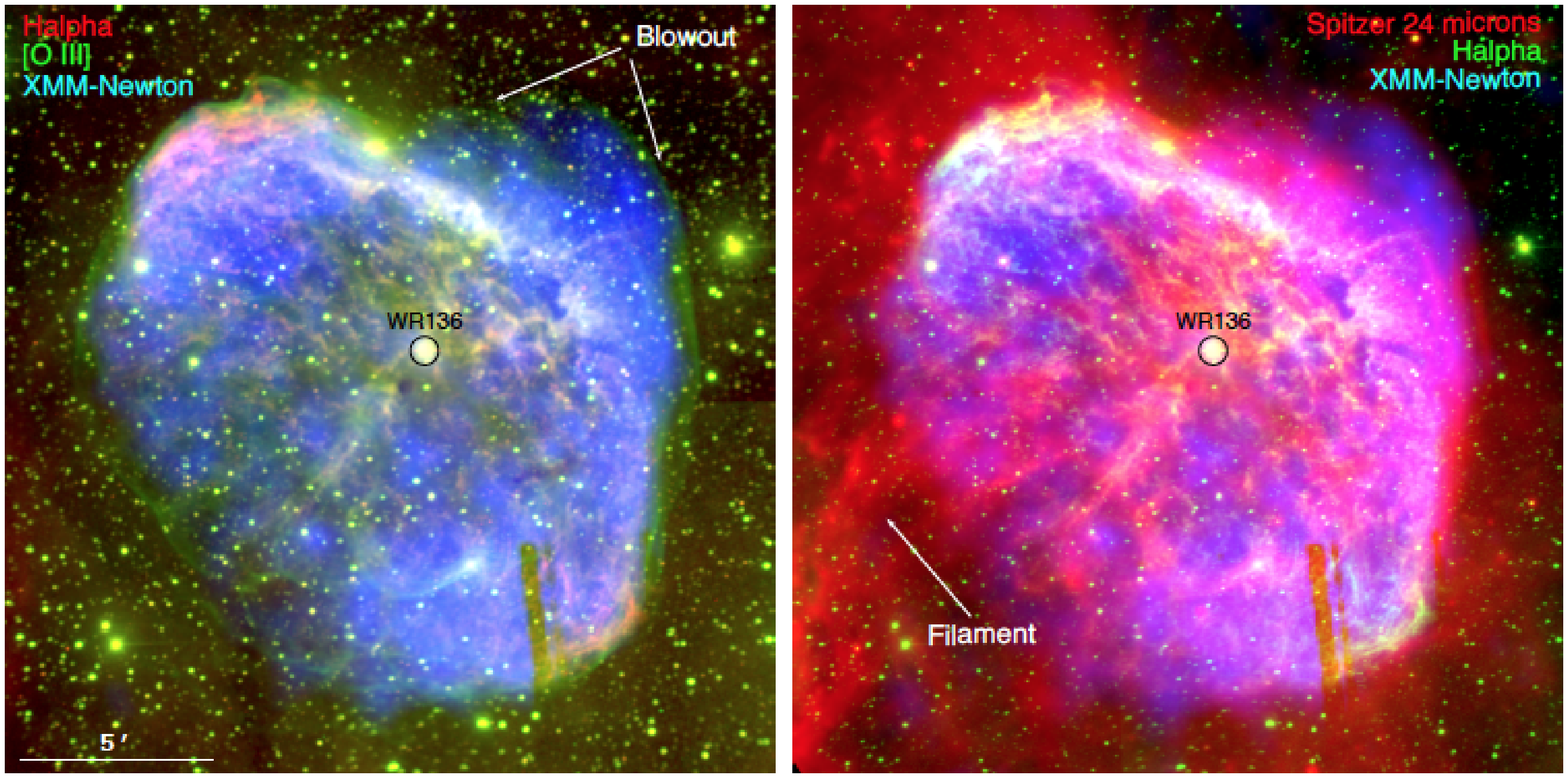}
\caption{
X-ray, optical, and mid-IR colour-composite pictures of NGC\,6888.  
Left: H$\alpha$ (red) and [O\,{\sc iii}] (green) narrowband images 
\citep{Gruendl2000}, and soft (0.3--0.7~keV) diffuse X-ray emission 
(blue). 
Right: {\it Spitzer} MIPS 24 $\mu$m (red), H$\alpha$ narrowband image 
(green), and soft (0.3--0.7~keV) diffuse X-ray emission (blue).  
The position of the central star, WR\,136, is shown in both panels. 
North is up, east to the left.  
}
\end{center}
\label{fig:RGB}
\end{figure*}

To further illustrate the distribution of the diffuse X-ray emission,
we have used the {\sc ciao} {\it dmfilth} routine \citep[{\sc ciao}
Version 4.6;][]{Fruscione2006} to excise all point-like sources and
create a clean view of the X-ray-emitting gas in the soft band.  The
final image is presented in Figure~2 in comparison with other
wavelengths.  Figure~2-left panel shows the comparison with the
H$\alpha$ and [O\,{\sc iii}] narrowband images obtained at the 1~m
telescope of the Mount Laguna Observatory \citep[MLO;
see][]{Gruendl2000}.  The X-ray-emitting gas is delimited by the
[O\,{\sc iii}] {\it skin} and not by the H$\alpha$ clumpy
distribution.  Figure~2-right panel shows the comparison of the
diffuse X-ray emission with that of the {\it Spitzer} MIPS 24~$\mu$m
and H$\alpha$ narrowband emission.  The mid-IR emission in the {\it
  Spitzer} MIPS 24~$\mu$m image also shows a skin at the location of
the northwest blowout that confines the diffuse soft X-ray emission.

\begin{figure}
\begin{center}
\includegraphics[angle=0,width=1.\linewidth]{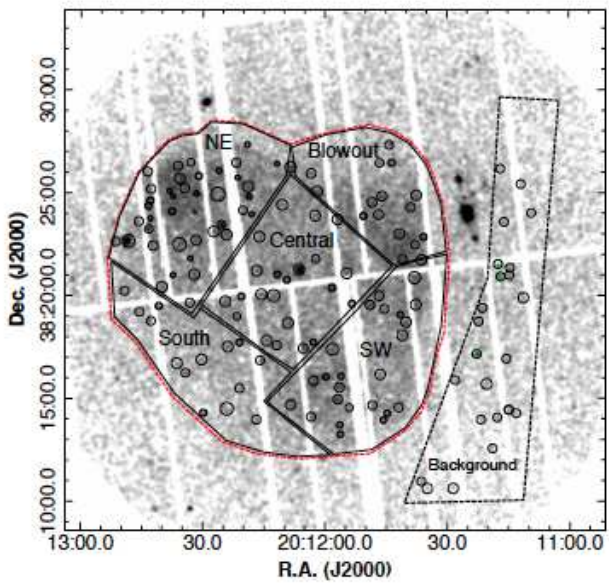}
\caption{
EPIC-pn event file of the field of view of NGC\,6888. 
The polygonal regions correspond to the source apertures used to 
extract the spectra, where the circular regions around point-like 
sources have been excised.  
The red line encompases the optical [O\,{\sc iii}] emission.  
The dashed-line polygon shows the region used for the selection of 
the background region.  
}
\end{center}
\label{fig:ngc6888_regiones}
\end{figure}

\section{PHYSICAL PROPERTIES OF THE HOT GAS IN NGC\,6888}

To proceed with the study of the physical properties of the
X-ray-emitting gas in NGC\,6888, we have extracted X-ray spectra from
different regions of this nebula. In order to identify periods of
high-background levels, we created light curves binning the data over
100~s for each of the EPIC cameras in the 10--12~keV energy band.
Background was considered high for count rate values over 1.0
counts~s$^{-1}$ and 0.2 counts~s$^{-1}$ for the pn and MOS cameras,
respectively.  The net exposure times after excising bad periods of
time are 48.8 ks, 63.8 ks, and 64.9 ks for the pn, MOS1, and MOS2
cameras respectively\footnote{ Note the differences in the net
  exposure times between the {\sc esas} tasks and this procedure
  \citep[see also][for the case of NGC\,2359]{Toala2015}.}.

In order to study the spectral variations within regions in NGC\,6888,
we have defined several polygonal apertures as shown in Figure~3. 
These apertures delineate specific nebular regions: the northeast (NE)
cap, the southwest (SW) cap, the blowout (B), the central region (C),
and the southern (S) region. An additional aperture, defined by a
source region (shown with red rashed-line in Fig.~3) encompassing the
[O\,{\sc iii}] line emission, correspond to the global diffuse X-ray
emission. The background has been extracted from a region close to the
camera edges that does not include diffuse X-ray emission from the
nebula.

The background-subtracted spectra for the different regions defined
above are shown in Figure~4. The top-left panel of Figure~4 shows the
spectra corresponding to the global X-ray emission in NGC\,6888
extracted from the three EPIC cameras (pn, MOS1, and MOS2). For the
apertures corresponding to specific nebular regions, we only show the
EPIC-pn spectra given the superior quality of these spectra as
compared to those extracted from the EPIC-MOS cameras.
The final EPIC-pn count rates and count number in the 0.3--1.5 keV for 
the different regions (NE, SW, blowout, central, and south regions) are 
presented in the first two columns of Table~1.

\subsection{Spectral Properties}
\label{sec:spectral_properties}

\begin{figure*}
\begin{center}
\includegraphics[angle=0,width=1.\linewidth]{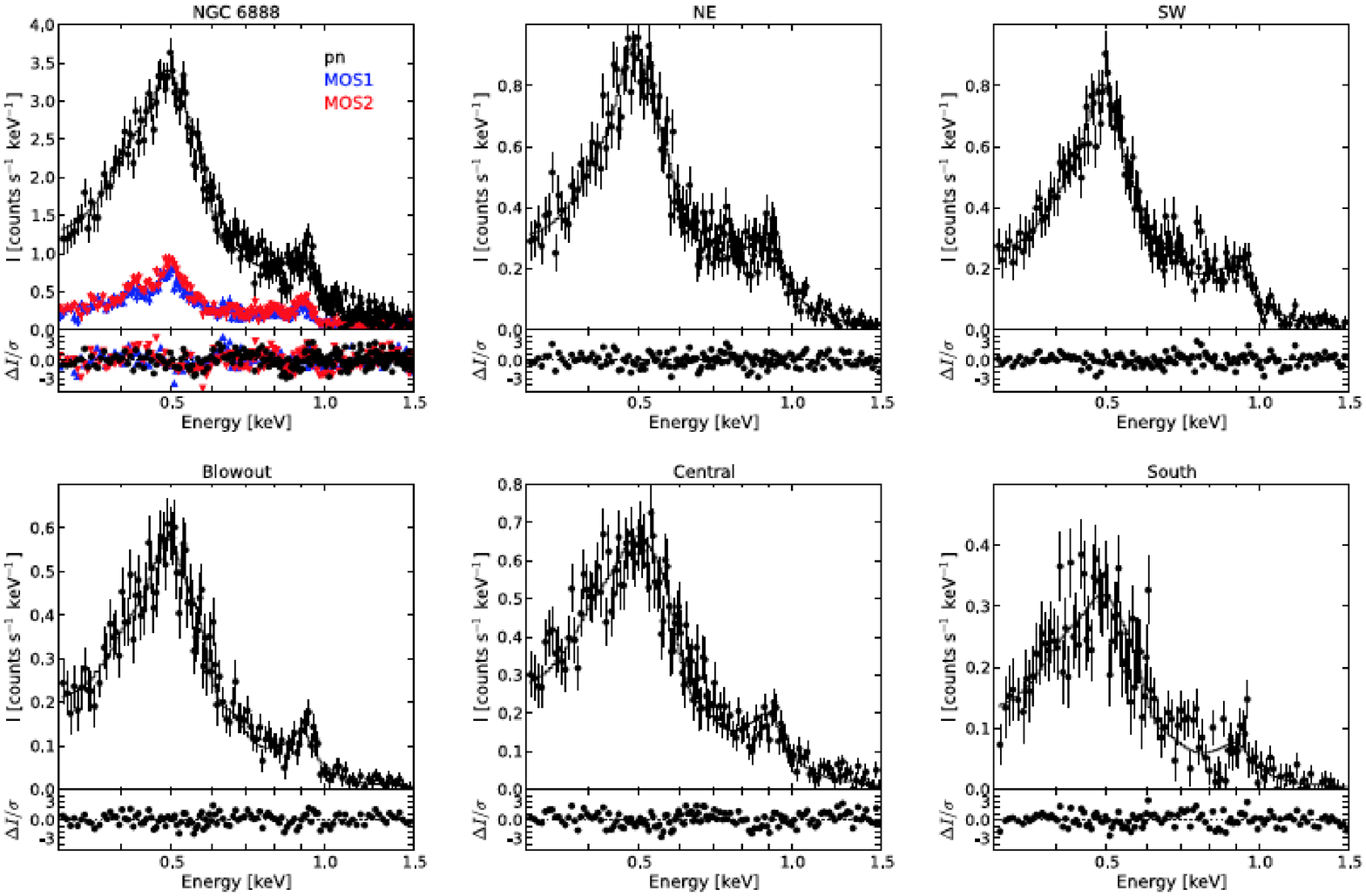}
\caption{
\emph{XMM-Newton} background-subtracted spectra of different regions of 
NGC\,6888 overplotted with their best-fit two-temperature \textit{apec} 
model shown by solid lines (top panels) and residuals of these fits 
(bottom panels).  
The EPIC pn, MOS1, and MOS2 spectra of the whole nebula (top-left) are 
displayed, whereas only the EPIC-pn spectra are shown for specific 
nebular regions.  
}
\label{fig:spec_EPIC_todo}
\end{center}
\end{figure*}

All spectra shown in Fig.~4 are soft and resemble those obtained from
\textit{Chandra} and {\it Suzaku} observations
\citep{Zhekov2011,Toala2014}.  The spectra exhibit a broad bump that
peaks at 0.5 keV, with a secondary peak at 0.7--1.0 keV.  No
significant emission is detected at energies above 1.5 keV. The main
spectral feature that peaks at 0.5 keV can be associated to the N~{\sc
  vii} 24.8 \AA\ emission line, while the secondary peak can be
attributed to the Fe complex and to Ne~{\sc ix} lines. The spectra in
Fig.~4 show subtle differences among them. For example, the 0.5~keV
feature of the spectra from the NE and SW regions seem rather narrow,
whilst the other regions (blowout, central, and south) show a broader
feature.

In order to assess these spectral differences we have calculated the
averaged ratio of intensities for different energies. For example, in
order to produce the ratio of intensities at 0.4 and 0.5 we have
averaged five spectrum channels around each enegy value. We have also
computed the FWHM of the emission feature centered at 0.5 keV,
FWHM($E_{0.5}$). This has been done for each spectra defining a basal
value which corresponds to the intensity at 0.8 keV. All these values
are presented in Table~1. Column~4 in Table~1 confirms that the
feature at 0.5 keV is narrower in the NE and SW spectra than in other
regions, with the blowout region having an intermediate
value. Furthermore, the ratio between the intensities at 0.4 and 0.5
keV, $R(0.4/0.5)$, shows that the contribution to the 0.4 keV is
notably larger in the central and southern regions. In particular, the
SW region has the narrowest 0.5 keV spectral feature and the smallest
contribution at 0.4 keV, with FWHM$(E_{0.5})$=0.14 keV and
$R(0.4/0.5)$=0.61.

\begin{table*}
\centering
\caption{Spectral features of the different regions defined in the EPIC-pn observations$^\mathrm{a}$.}
\begin{tabular}{lccccccc}
  \hline
  \hline
  Region & Count rate    &  Counts        &FWHM($E_{0.5}$) &  $R(0.4/0.5)^\mathrm{b}$ & $R(0.8/0.5)^\mathrm{b}$ & $R(0.9/0.5)^\mathrm{b}$ & $R(0.9/0.4)^\mathrm{b}$ \\ 
    & (counts~ks$^{-1}$) &                & (keV)          &               &              &              &              \\
  \hline
  NE     & 315           & 14800$\pm$500 &   0.15          & 0.64         & 0.29          & 0.34        & 0.53          \\
  SW     & 260           & 11900$\pm$200 &   0.14          & 0.61         & 0.27          & 0.27        & 0.43          \\
  Blowout& 180           & 8500$\pm$150  &   0.19          & 0.65         & 0.18          & 0.22        & 0.34          \\
  Central& 250           & 11600$\pm$170 &   0.22          & 0.83         & 0.26          & 0.28        & 0.33          \\
  South  & 110           & 5050$\pm$150  &   0.22          & 1.12         & 0.24          & 0.30        & 0.27          \\
  \hline
  \hline
\end{tabular}
\label{tab:table1}
\begin{list}{}{}
\item{$^\mathrm{a}$All values were obtained from EPIC-pn spectra.}
\item{$^\mathrm{b}$Typical error values are $<$0.01.}
\end{list} 
\end{table*}

To finally assure the differences between the background-subtracted
spectra, we have run Kolmogorov-Smirnof (KS) statistics tests using
the SciPy Python library. The KS statistic turned out to be the lowest
between the NE and SW spectra, with a value of 0.15. Other KS tests
turned out to be higher when comparing the NE (or SW) with spectra
from other regions. For example, the KS statistic resulted to be 0.47
when comparing the NE and south region. As lower KS statistics mean
that the two samples are more similar, these results prompt us to
anticipate that the physical properties between the NE and SW regions
will be similar than compared to other regions.
 
Three different scenarios can be envisaged to interpret these spectral
variations: i) nitrogen abundance variations in different regions, ii)
variations in the plasma temperatures, that is, if the caps would have
higher temperatures it would produce larges emissivity in the N\,{\sc
  vii} line, and iii) higher values of the extinction towards the NE
and SE caps would reduce the emission in the lowest energy range.

To study the physical properties of the X-ray emission, all spectra
shown in Figure~4 have been modeled and fit with XSPEC v12.8.2
\citep{Arnaud1996}, using an absorbed two-temperature \textit{apec}
optically thin plasma emission model with a \textit{tbabs} absorption
model as described in \citet{Wilms2000}. In accordance with previous
X-rays studies of S\,308, NGC\,2359, and NGC\,6888, we use the low
temperature component to model the bulk of the X-ray emission, while
the higher temperature component models the emission above
$\sim$0.7~keV \citep[][and references
therein]{Toala2012,Toala2014,Toala2015}. We initially adopted the same
nebular abundances as in \citet{Toala2014} for N, O, and Ne of 3.2,
0.41, and 0.85 times their solar values \citep{Anders1989}, which are
averaged values from abundances found by \citet{FernandezMartin2012}
and S abundance of 0.39 times its solar value as found by
\citet{Moore2000}. Models with variable C, Mg, and Fe were tested but
their fitted abundances converged to solar values, so we fixed these
abundances. Models with variable Ne were also tested and the final
best-fit values ranged around 0.80 times its solar value. Thus, we
kept its value at 0.85 as found by \citet{FernandezMartin2012}. We
note that the N/O ratio used here is consistent with those results
reported recently by \citet{MesaDelgado2014}, \citet{Stock2014}, and
\citet{ReyesPerez2015}.

To evaluate further the three differences scenarios that can cause the
spectral differences, we ran different sets of spectral fits varying
the nitrogen abundance ($X_\mathrm{N}$) and plasma temperatures
($kT_{1}$ and $kT_{2}$). Even though extinction has been found to be
uniform throughtout NGC\,6888 by \citet{Wendker1975} and implies a
column density of $N_{\mathrm{H}}$=3.13$\times$10$^{21}$~cm$^{-2}$
\citep[see][]{Hamann1994}, we also explore some models with varying
hydrogen column density.

The resultant model spectra were compared with the observed spectra in
the 0.3--1.5~keV energy range and the $\chi^2$ statistics was used to
determine the best-fit models. 
A minimum of 100~counts per bin was required for the spectral fit. 
The resultant best-fit model parameters for all regions are listed in 
Table~2: plasma temperatures ($kT_{1}$, $kT_{2}$), normalization 
parameters ($A_{1}$ and $A_{2}$)\footnote{
The normalization parameter is defined as $A$=1$\times10^{-14}\int
n_{\mathrm{e}} n_{\mathrm{H}} dV/4 \pi d^{2}$, where $d$ is the distance, 
$n_\mathrm{H}$ and $n_\mathrm{e}$ are the number hydrogen and electron 
density, respectively, and $V$ the volume in cgs units.
}, 
nitrogen abundance ($X_{\mathrm{N}}$), column density ($N_\mathrm{H}$),
absorbed and unabsorbed fluxes ($f$ and $F$) in the 0.3--1.5~keV 
energy range, and reduced $\chi^{2}$.
We describe these fits in the two following sections.

\subsection{Global Properties of the hot gas in NGC\,6888}
\label{sec:global}

To study the global physical properties of NGC\,6888 we have fitted
the pn, MOS1, and MOS2 spectra simultaneously.  For this, we set the
column density to a fixed value
($N_{\mathrm{H}}$=3.13$\times$10$^{21}$ cm$^{-2}$) and varied the
plasma temperatures and nitrogen abundance. The result of this joint
fit is given in Table~2 labeled as `NGC\,6888'.  The temperature of
the two plasma components are $T_{1}$=1.4$\times$10$^{6}$~K
($kT_{1}=0.120$~keV) and $T_{2}$=8.2$\times$10$^{7}$~K
($kT_{2}=0.71$~keV), for a nitrogen abundance $X_{\mathrm{N}}$=3.70
times its solar value.  The total absorbed flux is
$f_{\mathrm{X}}$=(2.5$\pm$0.1)$\times$10$^{-12}$~erg~cm$^{-2}$~s$^{-1}$,
which corresponds to an unabsorbed flux of
$F_{\mathrm{X}}$=(4.2$\pm$0.2)$\times$10$^{-11}$~erg~cm$^{-2}$~s$^{-1}$.
Adopting a distance of 1.26~kpc, the X-ray luminosity is
$L_{\mathrm{X}}$=(7.9$\pm$0.3)$\times$10$^{33}$~erg~s$^{-1}$.

In order to estimate the electron density of the X-ray-emitting gas we
can assume a spherical morphology \citep[see][for a discussion on
adopting different morphologies]{Toala2014}. At a distance of 1.26~kpc
the maximum extent of the nebula, that is 9\arcmin, represents a
physical radius of 3.3~pc. Using the definition of the normalization
parameter ($A$) we can estimate an electron density of
$n_\mathrm{e}=0.4(\epsilon/0.1)^{-1/2}$~cm$^{-3}$, which implies a
mass of $m_\mathrm{X}=1.7(\epsilon/0.1)^{1/2}$~M$_{\odot}$, with
$\epsilon$ as the filling factor. Note that these values are the same
as those estimated from the {\it Chandra} observations which only
covered $\sim$60\% of the nebula \citep[see][]{Toala2014}.

The reduced $\chi^2$ of the above fit is good, but not optimal (1.45).
This implies that there is an underlying complexity to the emitted
spectrum, exceeding the simple model here proposed, but it may point
also to calibration discrepancies among the three EPIC cameras.  To
investigate this issue, we also performed model fits to the data of
NGC\,6888 obtained solely from the EPIC-pn camera. The best-fit model
parameters to this emission are also shown in Table~2 labeled as
'EPIC-pn \#1'. The model delivered similar temperatures for the
plasma, $T_{1}$=1.4$\times$10$^{6}$~K and
$T_{2}$=8.6$\times$10$^{7}$~K, and a nitrogen abundance of
$X_{\mathrm{N}}$=3.75 times its solar value.  The corresponding total
absorbed and unabsorbed fluxes are
$f_{\mathrm{X}}$=(2.6$\pm$0.2)$\times$10$^{-12}$~erg~cm$^{-2}$~s$^{-1}$
and
$F_{\mathrm{X}}$=(4.1$\pm$0.3)$\times$10$^{-11}$~erg~cm$^{-2}$~s$^{-1}$.
The corresponding luminosity is
$L_{\mathrm{X}}$=(7.8$\pm$0.6)$\times$10$^{33}$~erg~s$^{-1}$. One more
model (\#2) was tested by fixing the nitrogen abundance
$X_\mathrm{N}$=4, but similar results were obtained (see Table~2). The
reduced $\chi^2$ of the fits to the EPIC-pn spectrum is somewhat
lowered (1.3), but still significantly above unity.

\subsection{Physical properties of the different regions}
\label{sec:physical_properties}

In order to study the variations of the physical conditions in the
plasma within NGC\,6888, we have modeled separately the different
spectra as extracted from different regions defined in Fig.~3 as NE,
SW, blowout, central, and south. We have tried three types of models:
i) in models labeled as \#1 only the column density has been set to
the fixed nominal value $N_\mathrm{H}=3.13\times$10$^{21}$ cm$^{-2}$,
ii) for models labeled as \#2 both the column density and the nitrogen
abundance have been fixed ($N_\mathrm{H}=3.13\times$10$^{21}$
cm$^{-2}$ and $X_\mathrm{N}=4$), finally iii) in models labeled as \#3
the nitrogen abundance and plasma temperatures have been fixed to
those values obtained from the global fit of NGC\,6888 using the three
EPIC cameras ($X_\mathrm{N}=4$, $kT_{1}$=0.120~keV, and
$kT_{2}$=0.71~keV), but the hydrogen column density has been set as a
free parameter of the model. The results are shown in Table~1.

The spectral fits obtained using the models mentioned above are
generally consistent with those found for the global spectra derived
from the three EPIC cameras and from the pn camera, although there is
evidence for variations in the plasma properties. The reduced $\chi^2$
are improved in the spectral fits, except for the central region. The
temperature of the main plasma component in the NE and SW regions
($T_{1}\sim$1.6$\times$10$^{6}$~K; models NE\#1, NE\#2, SW\#1, and
SW\#2) seems to be a bit higher than in other regions
($T_{1}\lesssim$1.4$\times$10$^{6}$ K; models B\#1, B\#2, C\#1, and
C\#2). On the other hand, the temperature of the second component
($T_{2}$=8--9$\times$10$^{6}$~K) is consistent among the different
regions, within the error bars. 
Models labeled as \#1 provide hints of enhanced nitrogen abundances in
the NE and SW regions, with nitrogen abundances somewhat higher
($X_\mathrm{N}\sim5$) than in other regions ($X_\mathrm{N}<3.9$).
Alternatively, models labeled as \#3 suggest that the NE and SW
spectra are more extinguished than the spectra extracted from other
regions. These detailed fits thus confirm the spectral differences
seen in Fig.~1 and revealed in Table~\ref{tab:table1}.

\section{DISCUSSION}

The WR nebula NGC\,6888 received attention from almost all previous
X-ray telescopes, such as \emph{Einstein}, \emph{ROSAT}, \emph{ASCA},
\emph{Suzaku}, and \emph{Chandra}, but this is the first time that it
has been mapped entirely at high-sensitivity with reasonable spatial
resolution ($\sim$6$^{\prime\prime}$). In general, the spatial
distribution of the X-ray emission resulting from the current
\emph{XMM-Newton} observations is in agreement with previous results,
although the X-ray images in Figures~1 and 2 reveal structures in
great detail. The morphology of the X-ray emission of NGC\,6888 is far
from simple: most emission comes from three maxima at the northeast
and southwest caps and northwest blowout, while the central and
southwestern regions show a rather homogeneous distribution of the
X-ray emitting gas. The X-ray emission from the northwest region is
confined by the northwest blowout delineated by optical [O~{\sc iii}]
and \emph{Spitzer} MIPS 24 $\mu$m emission.

Previous X-ray observations of NGC\,6888 did not have the capabilities
of resolving spectral differences across the nebula, consequently only
its global properties could be derived. The \emph{Chandra}
observations of NGC\,6888 presented by \citet{Toala2014} hinted at the
presence of spatial variations in the spectral properties of its
diffuse X-ray emission, but the differing spectral responses at low
energies of the two \emph{Chandra} ACIS-S CCDs that registered this
emission cast uncertainties on this finding. The unprecedented
sensitivity, full spatial coverage and homogeneous spectral response
of the current observations enabled the search for unambiguous spatial
variations in the spectral properties of the diffuse X-ray emission
from NGC\,6888. This resulted in the conclusive detection of spectral
differences at different locations of a WR nebula for the first time.
These are clearly revealed in Figures~1 and 4 and corroborated by
Tables~1 and 2.

In order to assess the origin of the spectral differences and explore
the three possible scenarios proposed in
Sec.~\ref{sec:spectral_properties}, we have performed different fits
to each of them labeled as \#1, \#2, and \#3 (see
Sec.~\ref{sec:physical_properties}). In particular, models labeled as
\#2 provided a clear support for variations of the main plasma
temperature, $kT_{1}$, of the caps with respect to the remaining
regions, from $(1.5^{+0.10}_{-0.06})\times$10$^{6}$~K in the caps down
to $(1.3^{+0.08}_{-0.07})\times$10$^{6}$~K in the central and southern
regions.  This trend is confirmed by the best-fit parameters of models
\#1, that suggest both nitrogen abundances enhancements, up to
$X_\mathrm{N}\sim5$, and larger plasma temperatures at the caps with
respect to the central and southernmost regions, with
$X_\mathrm{N}<3.9$. With models labeled as \#3 we wanted to assess the
possible variations of the absorption column density. The best-fit
parameters of the latter models suggest that the NE and SW regions
(the caps) have slightly larger $N_\mathrm{H}$, but the extinction
variations among the different regions, all within the error bars, are
not conclusive. Furthermore, if column density variations were the
reason for the differences in the spectra shown in
Fig.~\ref{fig:spec_EPIC_todo} we would have expected larger values
towards the central and south regions as suggested by the {\it
  Herschel} PACS 160~$\mu$m image of NGC\,6888 (Toal\'{a} et al. in
prep.). Finally, we would like to mention that in all different fits,
the blowout region shows intermediate values of plasma temperature and
nitrogen abundances.

The above spectral analysis leads us to conclude that the diffuse
X-ray emission from NGC\,6888 is not affected by significant
extinction variations, but rather it indicates correlated variations
in the plasma temperatures and nitrogen abundances. Regions with lower
plasma temperatures (i.e., the central and southern regions) have
nitrogen abundances closer to those of the optical nebula
\citep[e.g.,][]{ReyesPerez2015}, whereas regions with higher plasma
temperatures (i.e., the caps) have higher nitrogen abundances. These
correlations can be interpreted as evidence of low mixing efficiency
between the stellar wind and the nebular material in the caps. In
other words, the higher temperature of the NE and SW regions implies
less mixing with the outer nebular material.

It could be argued that the contrasting mixing efficiency across the
nebula is caused by the presence of a magnetic field which would
suppress thermal conduction in the direction perpendicular to the
magnetic lines. At this time, however, there is no definitive
detection of magnetic fields in WR\,136
\citep[e.g.,][]{Chevroti2014,Kholtygin2011} or its associated nebula
\citep{Wendker1975}. The physical structure of NGC\,6888 and its
surrouding medium offer alternative explanations. The nebula seems to
be located at the edge of a cold cloud revealed by its infrared
emission \citep[see][]{Gv2010}. This is illustrated by the
\emph{Spitzer} MIPS 24 $\mu$m emission from the nebula and its
surrouding medium shown in red in the right panel of Figure~2. The
spatial coincidence between the 24 $\mu$m emission from the nebula and
the H$\alpha$ emission from clumps and filaments and the [O\,{\sc
  iii}] emission from the blowout suggests a contribution of ionized
material to the nebular emission in this band and a contribution by
thermal dust emission \citep[e.g.,][]{Toala2015b}. The lack of mid-IR
emission towards the northwest of the WR bubble reveals the low
density of the ISM along this direction, naturally explaining the
production of the blowout.
On the other hand, the caps are formed as the stellar wind of the WR
star sweeps up material along directions of higher density.
The anisotropic of the ISM around WR\,136 can be expected to produce
different shock structures, depending on the density of the
surrounding material. These would result in different types of
hydrodynamical instabilities with varying efficiencies in the
injection and mixing of cold nebular material into the hot bubble
along different directions.

\subsection{Comparison with other Wolf-Rayet Nebulae}

Currently, there are three WR bubbles that have been reported to harbor 
diffuse X-ray emission, namely, S\,308, NGC\,2359, and NGC\,6888 around 
WR\,6, WR\,7, and WR\,136, respectively.  
The best-quality observations of these WR bubbles, those performed by 
\emph{XMM-Newton} \citep[][this work]{Chu2003,Toala2012,Toala2015}, 
suggest different scenarios for their formation.

The simplest case would be that of the WR nebula around WR\,6, S\,308.
The H\,{\sc i} distribution around the nebula \citep{Arnal1996}
indicates that the dense and slow wind material previous to the WR
phase was ejected into a low density cavity, probably carved by the
stellar feedback during the time the star was on the MS
phase. Moreover, radiative-hydrodynamic simulations require WR\,6 to
evolve through a yellow supergiant stage (YSG), with correspondingly
high dense wind velocities ~75~km~s$^{-1}$ \citep{Humphreys2010}, in
order to produce a WR bubble of radius as large as that of S\,308,
$\sim$9~pc \citep{Toala2011}. As a result, the stellar wind during the
subsequent WR phase swept smoothly the previous dense material into a
roundish morphology.

Meanwhile, the WR bubble NGC\,2359 around WR\,7 presents a more complex 
morphology, with a main bubble and several filaments and blisters 
\citep[see Fig.~1 and 3 in][]{Toala2015}.  
X-ray-emitting gas is detected towards the central cavity and inside the 
northeast blister, with no emission detected towards the southern blister 
due to high extinction at this region 
\citep[][and references therein]{Toala2015}.
These features are suggestive of the evolution of WR\,7 through 
eruptive and non-isotropic ejections of dense material in an LBV 
episode \citep[e.g.,][]{Rizzo2003}, but also reveal a complex CSM 
around WR\,7.

Finally, NGC\,6888 presents a clear case of the formation of a WR
bubble in a highly anysotropic CSM. The added complexity of the CSM
produces varying interactions of the fast stellar wind along different
directions resulting in anysotropic distributions of the plasma
temperature and abundances of the X-ray-emitting material.

The detection of X-ray emission inside these three WR bubbles and the
non-detections in RCW\,58 and the WR nebula around WR\,16
\citep{Gosset2005,Toala2013} are revealing.  Despite the different
scenarios for the formation of the three X-ray-emitting WR nebulae,
their central stars share similar spectral types and wind properties
\citep[WN4--6, $v_{\infty} \approx 1700$~km~s$^{-1}$, and $\dot{M}
\approx 5\times$10$^{-5}$ M$_{\odot}$~yr$^{-1}$; see][]{Hamann2006}.
These properties are at variance with those of the WR stars whose
wind-blown bubbles are not detected in X-rays: WN8h stars with
similarly high mass-loss rates, but lower stellar wind velocities
$v_{\infty}=650$ km~s$^{-1}$ \citep{Hamann2006}.

The current X-ray observations towards WR nebulae suggest that weak
stellar winds are not capable of producing diffuse X-ray emission.
The theory of wind-blown bubbles predicts that the current fast
stellar wind from the central star will ram onto the nebular material
producing an adiabatically shocked region, the so-called hot bubble,
with temperatures as high as 10$^{7}-10^{8}$ K
\citep[see][]{Dyson1997}. We remark that this theoretical prediction
is expected even for stellar wind velocities as low as 650
km~s$^{-1}$.  The low X-ray temperatures ($T \sim$10$^{6}$~K) implied
by the \emph{XMM-Newton} observations of WR bubbles suggest that the
hot plasma inside these bubbles results from the mixing between
shocked stellar wind and nebular material due to instabilities formed
in the wind-wind interaction zone and/or thermal conduction by hot
electrons \citep[e.g.,][and references
therein]{Freyer2006,Toala2011,Dwarkadas2013}.  Somehow, weak stellar
winds cannot enhance the mixing between the outer ionized material and
the shocked stellar wind and, thus, their WR nebulae do not exhibit
diffuse X-ray emission.  New physical processes (e.g., dust cooling,
anisotropic CSM, ...) need to be incorporated into theoretical
simulations to provide a realistic description of the observations.

\section{SUMMARY}


This work presents \emph{XMM-Newton} observations of the WR bubble
NGC\,6888 around WR\,136.  We have completed a high-spatial resolution
mapping of the nebula for the first time. This has allowed us to
corroborate the \citet{Toala2014} suggestion that the diffuse X-ray
emission fills the northwest blowout. The diffuse X-ray emission fills
the [O\,{\sc iii}] emission and is not only spatially related to the
inner H$\alpha$ clumpy distribution as suggested by previous authors.
The complex distribution of the X-ray-emitting gas within the WR
bubble, as well as the distribution of ionized gas and the outer
infrared emission seem to point out that the formation scenario of
NGC\,6888 is not as simple as that proposed by the classic wind-wind
interaction model
\citep[e.g.,][]{GS1996a,GS1996b,Freyer2006,Toala2011}.

The uniquely high count rate obtained from the present observations
allows a detailed study of the spectral variations across the nebula
for the first time. Five different regions have been identified in the
X-ray emission: two associated with the caps seen in optical and
infrared images and one with the blowout, one to the central region,
and one to the southern region. Our analysis suggests that the caps
have suffered less mixing of material from the nebula, having higher
temperatures and higher nitrogen abundance than those from other
regions.  The other regions exhibit lower plasma temperatures and
lower nitrogen abundances, closer to the nebular abundances, i.e., the
mixing has been more efficient.  Such variations could be due to the
presence of a magnetic field (not reported to date) or to different
shock patterns created by the interaction of the stellar wind of
WR\,136 with the inhomogeneous CSM around NGC\,6888.

The total unabsorbed X-ray flux in the 0.3--1.5~keV energy band is
estimated to be
$F_\mathrm{X}$=(4.2$\pm$0.2)$\times$10$^{-11}$~erg~cm$^{-2}$~s$^{-1}$
which corresponds to a luminosity of
$L_{\mathrm{X}}$=(7.9$\pm$0.3)$\times$10$^{33}$~erg~s$^{-1}$ at a
distance of 1.26~kpc. The rms electron density was estimated to be
$n_\mathrm{e}$=0.4$(\epsilon/0.1)^{-1/2}$ which results in a total
mass of the X-ray-emitting gas of
$m_\mathrm{X}=1.7(\epsilon/0.1)^{1/2}$~M$_{\odot}$, with $\epsilon$ as
the gas filling factor.

\section*{Acknowledgements}
\addcontentsline{toc}{section}{Acknowledgements}

This work was based on observations obtained with \emph{XMM-Newton}, 
an ESA science mission with instruments and contributions directly 
funded by ESA Member States and NASA. 

MAG and JAT are supported by the Spanish MICINN (Ministerio de Ciencia 
e Innovaci\'on) grants AYA 2011-29754-C03-02 and AYA 2014-57280-P. 
SJA and JAT acknowledge financial support through PAPIIT project 
IN101713 from DGAPA-UNAM (Mexico).


\begin{landscape}
\begin{center}
\begin{table}
\label{tab:spectral}
\centering
\caption{Spectral Fits of the Diffuse X-ray Emission from NGC\,6888$^{\star}$}
\begin{tabular}{lccccccccccccccl}
\hline
\hline
Region  & Label & $N_\mathrm{H}$ & $X_{\mathrm{N}}/X_{\mathrm{N},\odot}$ & $kT_{1}$      & $A_{1}^{\dagger}$  & $f_{1}^{\ddagger}$   & $F_{1}^{\ddagger}$   & $kT_{2}$  & $A_{2}^{\dagger}$    & $f_{2}^{\ddagger}$   &  $F_{2}^{\ddagger}$  &  $F_{1}/F_{2}$ & $T_{1}/T_{2}$ & $\chi^{2}$/DoF \\
        &       & ($\times$10$^{21}$~cm$^{-2}$) &   & (keV)                   &(cm$^{-5}$)        &(cgs)& (cgs) &(keV)            & (cm$^{-5}$)       &(cgs)& (cgs) & & \\      
\hline
NGC\,6888& \#1 & {\bf 3.13} &3.70$^{+0.23}_{-0.22}$& 0.120$^{+0.003}_{-0.002}$ & 5.0$\times10^{-2}$& 1.9$\times10^{-12}$& 4.1$\times10^{-11}$ & 0.71$^{+0.02}_{-0.02}$ & 6.1$\times10^{-4}$ & 5.9$\times10^{-13}$ & 1.8$\times10^{-12}$  &  22.2 & \dots & 1.45=874.88/600 \\
\hline
EPIC-pn  & \#1 & {\bf 3.13} & 3.75$^{+0.35}_{-0.32}$& 0.122$^{+0.004}_{-0.003}$ & 4.6$\times10^{-2}$& 2.0$\times10^{-12}$& 3.9$\times10^{-11}$ & 0.74$^{+0.03}_{-0.03}$ & 6.0$\times10^{-4}$ & 5.8$\times10^{-13}$ & 1.7$\times10^{-12}$  &  22.4 & \dots & 1.30=307.24/241 \\
         & \#2 & {\bf 3.13} & {\bf 4.0}           & 0.123$^{+0.003}_{-0.003}$ & 4.3$\times10^{-2}$& 2.0$\times10^{-12}$& 3.8$\times10^{-11}$ & 0.74$^{+0.03}_{-0.03}$ & 6.1$\times10^{-4}$ & 5.9$\times10^{-13}$ & 1.8$\times10^{-12}$  &  21.1 & \dots & 1.32=314.08/237 \\
\hline
NE       & \#1 & {\bf 3.13} & 4.72$^{+0.71}_{-0.95}$& 0.139$^{+0.004}_{-0.014}$ & 7.9$\times10^{-3}$& 5.6$\times10^{-13}$& 9.3$\times10^{-12}$ & 0.73$^{+0.05}_{-0.04}$ & 1.8$\times10^{-5}$ & 1.8$\times10^{-13}$ & 5.5$\times10^{-13}$  &  16.8 & 0.19 & 1.14=214.35/188 \\
         & \#2 & {\bf 3.13} & {\bf 4.0}                 & 0.129$^{+0.007}_{-0.005}$ & 1.0$\times10^{-2}$& 5.3$\times10^{-13}$& 9.8$\times10^{-12}$ & 0.71$^{+0.03}_{-0.05}$ & 2.0$\times10^{-4}$ & 1.9$\times10^{-13}$ & 5.9$\times10^{-13}$  &  16.7 & 0.18 & 1.15=187.60/163 \\
         & \#3 & 3.70$^{+0.27}_{-0.26}$ & {\bf 4.0} & {\bf 0.120} & 1.9$\times$10$^{-2}$ & 5.4$\times$10$^{-13}$ &  1.7$\times$10$^{-11}$ & {\bf 0.71} & 2.2$\times$10$^{-4}$  & 1.8$\times$10$^{-13}$ & 8.0$\times$10$^{-13}$ & 20.3 & 0.17 & 1.14=187.16/164 \\
\hline
SW       & \#1 & {\bf 3.13} & 5.50$^{+1.00}_{-0.80}$& 0.143$^{+0.005}_{-0.004}$ & 6.8$\times10^{-3}$& 5.6$\times10^{-13}$& 8.9$\times10^{-12}$ & 0.78$^{+0.08}_{-0.07}$ & 1.3$\times10^{-4}$ & 1.3$\times10^{-13}$ & 3.8$\times10^{-13}$  &  22.9 & 0.18 & 0.97=178.40/184 \\
         & \#2 & {\bf 3.13} & {\bf 4.0}                 & 0.130$^{+0.009}_{-0.006}$ & 1.0$\times10^{-2}$& 5.4$\times10^{-13}$& 9.9$\times10^{-12}$ & 0.72$^{+0.04}_{-0.06}$ & 1.7$\times10^{-4}$ & 1.6$\times10^{-13}$ & 4.9$\times10^{-13}$  &  20.1 & 0.18 & 1.03=149.75/145 \\
         & \#3 & 3.85$^{+0.32}_{-0.31}$ & {\bf 4.0} & {\bf 0.120} & 2.3$\times$10$^{-2}$ & 5.9$\times$10$^{-13}$ &  1.9$\times$10$^{-11}$ & {\bf 0.71} & 1.8$\times$10$^{-4}$  & 1.1$\times$10$^{-13}$ & 7.0$\times$10$^{-13}$ & 27.1 & 0.17 & 1.00=144.62/146 \\
\hline
Blowout\,(B)  & \#1 & {\bf 3.13} & 3.85$^{+0.65}_{-0.55}$& 0.125$^{+0.008}_{-0.006}$ & 7.6$\times10^{-3}$& 3.5$\times10^{-13}$& 6.8$\times10^{-12}$ & 0.80$^{+0.09}_{-0.08}$ & 6.3$\times10^{-5}$ & 6.3$\times10^{-14}$ & 1.8$\times10^{-13}$  &  37.4 & 0.16 & 0.98=117.16/119 \\
         & \#2 & {\bf 3.13} & {\bf 4.0}                 & 0.126$^{+0.007}_{-0.005}$ & 7.3$\times10^{-3}$& 3.5$\times10^{-13}$& 6.7$\times10^{-12}$ & 0.80$^{+0.09}_{-0.08}$ & 6.4$\times10^{-5}$ & 6.3$\times10^{-14}$ & 1.8$\times10^{-13}$  &  36.7 & 0.15 & 0.98=117.31/120 \\
         & \#3 & 3.4$^{+0.32}_{-0.31}$ & {\bf 4.0} & {\bf 0.120} & 1.1$\times$10$^{-2}$ & 3.6$\times$10$^{-13}$ &  9.4$\times$10$^{-12}$ & {\bf 0.71} & 6.3$\times$10$^{-5}$  & 5.4$\times$10$^{-14}$ & 1.8$\times$10$^{-13}$ & 52.3 & 0.17 & 0.98=118.88/121 \\
\hline
Central\,(C) & \#1 & {\bf 3.13} & 2.90$^{+0.51}_{-0.50}$& 0.115$^{+0.005}_{-0.004}$ & 1.2$\times10^{-2}$& 3.7$\times10^{-13}$& 8.2$\times10^{-12}$ & 0.78$^{+0.05}_{-0.05}$ & 1.1$\times10^{-4}$ & 1.0$\times10^{-13}$ & 3.0$\times10^{-13}$  &  27.0 & 0.15 & 1.35=185.31/137 \\
         & \#2 & {\bf 3.13} & {\bf 4.0}                 & 0.121$^{+0.006}_{-0.004}$ & 8.5$\times10^{-3}$& 3.6$\times10^{-13}$& 7.3$\times10^{-12}$ & 0.77$^{+0.05}_{-0.05}$ & 1.0$\times10^{-4}$ & 1.0$\times10^{-13}$ & 3.1$\times10^{-13}$  &  23.7 & 0.15 & 1.42=196.62/138 \\
         & \#3 &  3.04$^{+0.32}_{-0.29}$ & {\bf 4.0} & {\bf 0.120} & 8.1$\times$10$^{-3}$ & 3.6$\times$10$^{-13}$ &  6.8$\times$10$^{-12}$ & {\bf 0.71} & 1.0$\times$10$^{-4}$  & 1.0$\times$10$^{-13}$ & 3.2$\times$10$^{-13}$ & 21.3 & 0.17 & 1.45=201.93/139 \\
\hline
South\,(S)   & \#1 & {\bf 3.13} & 3.71$^{+1.20}_{-0.90}$& 0.113$^{+0.008}_{-0.008}$ & 5.7$\times10^{-3}$& 1.9$\times10^{-13}$& 4.2$\times10^{-12}$ & 0.79$^{+0.13}_{-0.14}$ & 4.2$\times10^{-5}$ & 4.1$\times10^{-14}$ & 1.2$\times10^{-13}$  &  35.0 & 0.14 & 1.17=186.76/159 \\
         & \#2 & {\bf 3.13} & {\bf 4.0}                & 0.115$^{+0.007}_{-0.006}$ & 5.2$\times10^{-3}$& 1.8$\times10^{-13}$& 4.1$\times10^{-12}$ & 0.78$^{+0.13}_{-0.15}$ & 4.3$\times10^{-5}$ & 4.2$\times10^{-14}$ & 1.2$\times10^{-13}$  &  33.2 & 0.15 & 1.23=145.33/118 \\
         & \#3 &  2.71$^{+0.53}_{-0.46}$ & {\bf 4.0} & {\bf 0.120} & 3.2$\times$10$^{-3}$ & 1.8$\times$10$^{-13}$ &  2.7$\times$10$^{-12}$ & {\bf 0.71} & 3.9$\times$10$^{-5}$  & 4.4$\times$10$^{-14}$ & 1.5$\times$10$^{-13}$ & 18.0 & 0.17 & 1.22=145.68/119 \\
\hline
\hline
\end{tabular}
\begin{list}{}{}
\item{$^{\star}$Boldface text represent fixed values on each fit.}
\item{$^{\dagger}$The normalization parameter is defined as $A=1
    \times 10^{-14} \int n_\mathrm{e} n_\mathrm{H} dV/ 4 \pi d^{2}$,
    where $d$, $n_\mathrm{e}$, $n_\mathrm{H}$, and $V$ are the
    distance, electron and hydrogen number densities, and volume in
    cgs units, respectively.}
\item{$^{\ddagger}$Fluxes ($f$ and $F$) are computed in the 0.3-1.5~keV energy range and are presented in cgs units (erg~cm$^{-2}$~s$^{-1}$).}
\end{list}
\end{table}
\end{center}
\end{landscape}

\end{document}